# The acute:chronic workload ratio: challenges and prospects for improvement

Chinchin Wang[1,2], Jorge Trejo Vargas[3], Tyrel Stokes[3], Russell Steele[3], Ian Shrier[1]

[1] Centre for Clinical Epidemiology, Lady Davis Institute, Jewish General Hospital, McGill University, 3755 Côte Ste-Catherine Road, Montreal, Quebec, Canada H3T 1E2

[2] Department of Epidemiology, Biostatistics and Occupational Health, McGill University, 1020 Pine Avenue West, Montreal, Quebec, Canada H3A 1A2

[3] Department of Mathematics and Statistics, McGill University, 805 Sherbrooke Street West, Montreal, Quebec, Canada H3A 0B9

**Corresponding Author:**

Ian Shrier MD, PhD

Centre for Clinical Epidemiology, Lady Davis Institute, Jewish General Hospital, McGill University, 3755 Côte Ste-Catherine Road, Montreal, Quebec, Canada H3T 1E2

Email: ian.shrier@mcgill.ca

Phone Number: 1-514-229-0114

**ORCIDs of authors:**

Chinchin Wang: https://orcid.org/0000-0002-8472-9936

Jorge Trejo Vargas: N/A

Tyrel Stokes: https://orcid.org/0000-0001-9305-1859

Russell Steele: https://orcid.org/0000-0003-1071-0914

Ian Shrier: https://orcid.org/0000-0001-9914-3498

## Abstract
Injuries occur when an athlete performs a greater amount of activity (workload) than what their body can absorb. To maximize the positive effects of training while avoiding injuries, athletes and coaches need to determine safe workload levels. The International Olympic Committee has recommended using the acute:chronic workload ratio (ACRatio) to monitor injury risk, and has provided thresholds to minimize risk. However, there are several limitations to the ACRatio which may impact the validity of current recommendations. In this review, we discuss previously published and novel challenges with the ACRatio, and possible strategies to address them. These challenges include 1) formulating the ACRatio as a proportion rather than a measure of change, 2) its use of unweighted averages to measure activity loads, 3) inapplicability of the ACRatio to sports where athletes taper their activity, 4) discretization of the ACRatio prior to model selection, 5) the establishment of the model using sparse data, 6) potential bias in the ACRatio of injured athletes, 7) unmeasured confounding, and 8) application of the ACRatio to subsequent injuries.

## Key Points
- The acute:chronic workload ratio is a measure recommended for use by the International Olympic Committee to guide training and minimize injury risk. However, there are several limitations associated with it that need to be addressed.
- The acute:chronic workload ratio should exclude the acute load from the chronic load calculation, and models should be developed using continuous ratios and larger sample sizes.
- Although the exponentially weighted moving average is appropriate in certain contexts, it suffers from an initial value problem and is not applicable to all sports.
- New analytical strategies must be developed to prevent bias and confounding in the acute:chronic workload ratio, and to apply it to recurrent injuries.

# 1. Introduction

Physical activity is widely regarded as an important part of a healthy lifestyle [1]. However, increased physical activity is associated with increased risk of injury, which may result in morbidity and time lost from activity [2]. Quantifying the relationship between physical activity and injury risk to determine a safe level of activity should improve the well-being of athletes and the general population.

Although muscle and tendon increase their strength in response to load [3], large increases in load that surpass their strength will cause damage. The unwritten rule of thumb for many years was that injury should be minimized if activity were increased by less than 10-20% per week. Recently, there has been increased interest in using changes in physical activity patterns to predict injury risk. Hulin et al. introduced the concept of an acute:chronic workload ratio (ACRatio) to measure changes in activity [4]. The ACRatio is calculated as the acute (i.e. recent) workload or activity divided by the chronic (i.e. long term) activity [4]. Although the acute workload is generally defined as the activity performed in the past week, and the chronic workload is generally defined as a weekly average of activity in the past 4 weeks, the optimal choice of time windows differ depending on the sport and training schedule [5]. The unit of workload depends on the context; it may be quantified as distance travelled, balls bowled, weight lifted, or session-rating of perceived exertion (RPE), among others [6].

In studies of cricket, rugby, Australian football, and soccer players, increases in ACRatio were associated with increases in injury risk [4,7–10]. A general model for the relationship between one's ACRatio and their risk of injury was generated using data from cricket, rugby, and Australian football players (Figure 1) [11]. It identified ratios in the "sweet spot" between 0.8 and 1.3 as being associated with the lowest risk of injury. Injury risks increased with decreasing ratios below 0.8, and increasing ratios above 1.3 [12]. These results supported the rule of thumb for a 10-20% increase in workload per week. This model was included in a consensus statement by the International Olympic Committee (IOC) and used as a guideline for training practices [12].

Although there is evidence to suggest that the ACRatio is a useful predictor of injury risk, some authors have noted limitations with its most common formulation and the model associated with it. The purpose of this article is to provide an overview of previously published and new unpublished challenges of the common use of the ACRatio. We also provide a summary of the type of data that would be necessary to avoid some of the more serious limitations associated with current analytical strategies, as well as additional novel analyses not previously suggested.

## 2. Limitations of the ACRatio

### 2.1 The ACRatio is a measure of the proportion of activity rather than the change

The ACRatio was intended to be used as a preventive tool for indicating whether an athlete has sufficiently prepared for their upcoming (acute) training load by comparing it to their previous (chronic) load [7]. If an athlete has a low acute load relative to their chronic load, they will be decreasing their relative activity and will have a low ACRatio. Athletes with low ACRatios are thought to be performing within the limits of what they have prepared for, and are therefore expected to have lower risks of injury. Conversely, if an athlete has a high acute load relative to their chronic load, they will have a high ACRatio. These athletes are expected to exceed the level of training they have prepared for, resulting in an increased risk of injury [6].

Although the ACRatio acts as a marker of athlete preparedness based on changes in activity [6], the conventional measure is actually a proportion rather than a true measure of change. The activity performed as part of the acute load is included in the calculation of the chronic load, making it a "mathematically coupled" measure [13]. The use of an "uncoupled" measure, where the acute load is excluded from the chronic load calculation, has also been proposed [13]. In this section, we will discuss the limitations associated with the coupled measure, and how they are resolved with the uncoupled measure.

There are three important limitations to the conventional coupled approach. First, it creates a spurious correlation between the acute and chronic loads [14]. This creates an apparent decrease in the variability of activity load between athletes, potentially leading to inappropriate inferences [14]. Using an uncoupled measure removes the correlation and increases the between-athlete variability [14].

Second, it creates a theoretical maximum ACRatio of 4, regardless of the magnitude of the absolute change in workload [13]. For example, consider an individual who has planned activity for the upcoming week, but was inactive in the three weeks prior. The numerator for their ACRatio will be the workload in that week (WL1). The denominator will be the average of the total workload from the 4 weeks, which is $0 + 0 + 0 + WL1$ (i.e. WL1/4). The ACRatio is therefore

$$\frac{WL1}{(\frac{WL1}{4})} = 4$$

no matter the magnitude of WL1. Therefore, the coupled ACRatio is unlikely to hold as a good predictor of injury risk at relatively high acute loads. The uncoupled measure excludes the upcoming week from the chronic load calculation and has no such maximum bound [13].

Therefore, one would expect it to better differentiate changes in workload that have different absolute magnitudes.

Third, the proportional nature of the ACRatio makes it difficult for coaches and athletes to calculate what an appropriate workload would be for the upcoming week. Determining an optimal acute training load using the coupled ACRatio requires consideration of how the chronic load will be affected by activity performed in the upcoming week that has not yet occurred. The uncoupled measure is simpler to calculate for athletes and coaches because one can simply divide the anticipated training in the upcoming week by their activity in the previous 4 weeks, without having to account for next week's activity in the chronic load calculation (Appendix 1).

The uncoupled measure solves the three limitations above, and follows the same patterns as the conventional calculation where large increases in acute load relative to the chronic load are associated with increased injury risk [15]. Although some have suggested that each individual should simply decide which method to use since it is possible to convert the coupled measure to the uncoupled one and vice versa [13], the "sweet spot" thresholds will be different for the two calculations which could create confusion. We believe the use of a single model and threshold is more practical to clinicians and athletes. Because the uncoupled measure resolves the above limitations of the coupled measure without introducing new limitations, we strongly recommend that the uncoupled measure be used to guide training recommendations.

## 2.2 The ACRatio uses unweighted averages to calculate workload

In the original formulation of the ACRatio, the chronic load is defined as the weekly average of the workload performed in the past 4 weeks [4]. Each of the 4 weeks is weighted equally in the calculation. While straightforward to calculate, the equal weighting obscures weekly variations in activity load and neglects that the effects of training decay over time [16]. Menaspà illustrated how 3 different athletes with vastly different training patterns can have the same ACRatio (Figure 2) [17]. Although each athlete would be assigned the same injury risk based on their shared ACRatio, their true injury risks would conceivably differ depending on how their workload was spread over the 4 weeks.

Similar problems arise with using a single weekly value for the acute load. Two athletes with the same acute workload may have different daily variations in activity that affect their risk of injury. For example, an athlete who performs all of his or her workload on a single day might be at higher risk of injury than an athlete who performs the same amount of workload spread over 7 days.

The underlying issue is that an injury is typically more likely to be associated with one's most recent workload than workload performed weeks prior. To solve this problem, some authors suggested that average workloads be calculated by providing more weight to recent workloads and less weight to workloads in the distant past [17,18]. Williams et al. proposed a weighting method that uses exponentially weighted moving averages (EWMA) for the calculation of chronic and acute loads [18]. The EWMA applies exponentially decreasing weights for workloads performed on each prior day, such that workloads performed on the first day of the chronic load calculation (i.e. the day furthest away from the day of calculation) and the first day of the acute load calculation should theoretically be weighted the least. Therefore, those who engage in greater recent activity will have higher loads, which may affect whether or not they are in a 'sweet spot' for minimizing injury risk.

Using EWMA weighting may better reflect changes in activity patterns and their effects on injury. EWMA-weighted ACRatios were more sensitive to changes in injury risk than traditional ACRatios in a cohort of 59 Australian football players, with an increase in $R^2$ from 0.21 in the traditional unweighted method to 0.87 using weighted averages [19]. Despite the results of these studies, there are several important limitations to this method that have not yet been reported.

The purpose of EWMA is to give more recent activity more weight, and activity further in the past less weight. However, mathematically, the initial load that is furthest away from the day of calculation may take on weight that is much higher than it should be, and in some cases even larger than the most recent workload. The EWMA calculation uses decay constants to apply decreasing weights to activity further in the past. When the decay constant is less than 0.5, the value at time 0 (most distant day from the day of calculation) takes on a greater weight than the subsequent days (Appendix 2). The difference between the weights at time 0 and the subsequent day decreases as the decay constant gets closer to zero (Appendix 3). This problem occurs with the proposed decay constants by Williams et al. (0.25 for the acute load and 0.068 for the chronic load [18]) when used to calculate loads for 28 days of activity. Although the initial load takes on a larger weight than subsequent days for both acute and chronic calculations, the decay constant is relatively large for the acute load, resulting in a fast rate of decay and a negligible difference between weights. The chronic load decay constant is much closer to zero, resulting in a large contribution from the initial load towards the weighted average (Figure 3). In fact, the contribution of the initial load is 1.9 times the contribution of the load on day 28 (the day of calculation) to the weighted average, even though the initial load should contribute the least weight.

One may choose to assign the first recorded load to the initial value [19], or use an arbitrary value, with 0 being an extreme case. Either value may take on an inappropriately large weight in the EWMA. However, with enough days included in the calculation, the EWMAs for an athlete with a positive initial load and with an initial load of 0 will converge. We illustrate this concept with an example using training loads from athlete 3 in Menaspà's paper [17]. We compare two scenarios – one where the initial load is 55, which is the first recorded load for athlete 3, and one where the initial load is 0 and the second load is 55. If this athlete repeats the same training profile 3 times consecutively, the two ACRatios will converge after around 50 days (Figure 4). This means that at least 50 days of activity is needed to reach convergence and avoid the initial value problem. In other words, we would have to ignore the first 50 days worth of information in order to obtain a valid estimate of injury risk, and this would only be valid for people who did not get injured during the first 50 days of training. Therefore, the methods of calculation for the decay constant or the EWMA itself may need to be changed.

Interpreting the results of the EWMA method presents two additional challenges. First, the acute load is no longer the activity in the most recent week, but uses the entire 4-week (28 day) period in its calculation. The weight given to days further than 7 days in the past is insignificant, but still present. The chronic load is similarly calculated using workloads from the entire 4-week period, but with different weights. Therefore, the EWMA remains a coupled measure. It may be more appropriate to use an uncoupled method whereby the EWMA is used to calculate an acute load from the past 7 days, and a chronic load from the 28 days prior to the acute load. Second, its complex calculation makes it difficult for athletes and coaches to determine the effect of changing activity patterns on their ACRatio. Rather, they must use a computer to determine whether their planned activity would result in an ACRatio that exceeds a particular risk threshold. This might be considered insignificant for elite athletes, but could represent a challenge for recreational and regionally competitive athletes who also want to maximize training and minimize injury risk.

## 2.3 The ACRatio does not account for tapering before competitions

The EWMA attempts to resolve the problem that activity performed in the distant past is less likely to cause injury than recent activity. Yet, this is not the case for all sports. Previous studies that demonstrated the utility of the EWMA looked at sports with competitive seasons [19]. However, when athletes train for a marathon, they typically taper their training load immediately prior to a race to recover from prior heavy training [20]. Tapering aims to reduce fatigue and increase recovery from training stress [21], thereby reducing risk of injury [22]. Similar training techniques are used by athletes who partake in sports with competitions lasting a short period of time, such as swimming, cycling, triathlon, and strength training [21]. In this context, tapering activity the few weeks before the competition decreases the chronic workload and thereby increases the ACRatio during the competition. If rest prior to increased activity during competition reduces injury risk as expected, the ACRatio will be a poor predictor of risk. Therefore, a modified model and set of recommendations should be investigated for these contexts.

## 2.4 The ACRatio is a continuous measure that has been discretized

The ACRatio can take on different values depending on how workload is measured. Workloads can be quantified in a variety of ways, including distance travelled, weight lifted, and session-RPE [6]. Regardless of the method of measurement and whether it is truly continuous, the ACRatio is essentially treated as a continuous measure with units that can conceivably take on any positive value in the uncoupled measure, or any positive value up to 4 in the coupled measure.

The majority of studies utilizing the ACRatio as a predictor of injury risk have categorized workloads and ratios into bins rather than treating the ACRatio as a continuous measure [4,5,11]. There are several benefits to the categorization of the ACRatio. Categorization of workloads into discrete ranges can aid athletes, coaches, and clinicians in formulating a training plan. Rather than calculating an exact amount of activity to perform to minimize injury, athletes can determine whether they can safely engage in low, moderate, or high amounts of activity based on their chronic load, offering them more flexibility in their training and easing interpretability. Discretization of a continuous measure such as the ACRatio can also increase efficiency and interpretability when developing models [23].

However, discretization prior to model selection in an analysis, as is typically done with the ACRatio, can result in a loss of information and hide the true relationship between variables [24]. Furthermore, different studies have binned ACRatios differently and used different reference levels [4,7,25,26], making it difficult to extrapolate findings and create a generalized model for injury risk. In simulations of training load and injury risk, discretization of the ACRatio resulted in increased false discovery rates when compared to continuous models [27]. Continuous models also provided more accurate estimated effects [27].

One reason for these findings may be related to sparse data bias. When the exposure is a categorical variable (e.g. discretized ACRatio) and the outcome is non-linear (e.g. dichotomous outcome such as injury), one generally requires 5 events for each level of each combination of covariates to avoid bias [28]. For example, categorizing the ACRatio into low-, medium-, and high-risk creates 3 levels. In a univariate analysis, one would need at least 15 events (i.e. injuries), 5 of which should occur in the low-risk group. This requirement is independent of the number of participants in each group. If one adjusts for additional variables like sex and age, 5 events would be required in each of the level combinations (e.g. 5 events in females aged 20-25 with a low ACRatio, 5 events in males aged 20-25 with a low ACRatio, etc.). If the ACRatio is continuous, one could instead fit splines, minimizing the challenges associated with sparse data [24]. Therefore, instead of discretizing before modeling, a more appropriate solution would be to use continuous data to develop the ACRatio-injury risk model and categorize the ACRatios into different risk levels after analysis. This would maintain interpretability while minimizing potential loss of data and bias.

## 2.5 The established ACRatio model is based on sparse data

Studies that have used the ACRatio to predict injury risk have been limited by small sample sizes, likely due to the difficulties of following a large group of participants over a time period that is long enough to result in a sufficient number of injuries. Nielsen et al. noted that out of 35 studies examining the relationship between training loads and injury risk, only 11 had a sample size greater than 150 [29]. These small studies create additional challenges beyond the sparse data bias mentioned above.

The pitfalls of having a small sample size are exemplified when the established model from the IOC consensus statement (Figure 1) is examined in greater detail. The model is based on sparse data derived from 3 studies with sample sizes of 28, 53, and an unreported number of athletes [4,7]. The model is based on discretized ACRatios, likely partially due to the relative lack of data at each value. ACRatios at the end-points (0 to 0.5 or 2.0+) were discretized as 0.5 or 2.0 respectively [11,12]. According to the model, the sweet spot for minimal injury risk is between 0.8 and 1.3, with a "danger zone" upwards of 1.5 [6]. However, the increase in injury risk appears to be driven completely by points at ACRatio = 2.0, which also differ substantially from each other in their corresponding likelihood of injury. Limiting the data to ACRatios lower than 2.0 removes any apparent relationship between ACRatio and injury risk.

Although this could indicate a "threshold" effect, the discretization of the endpoint data makes it impossible to determine whether there is a threshold, and if so, whether the increase in risk occurs at ACRatios of 2.0 or at higher values. This has implications for athletes trying to maximize their upcoming activity load while minimizing injury risk. If the true threshold for injury is higher than 2.0, then athletes could increase their acute load to a higher relative amount without increasing their risk of injury. Therefore, even though the IOC consensus statement supports the unwritten rule of thumb that a 10-20% increase in activity per week is safe, there remains limited evidence for its recommendations on current activity thresholds and sweet spots. Modelling training loads and injury risk with larger sample sizes and using continuous models will provide more accurate thresholds and training recommendations. Once the appropriate models are determined, investigators can then discretize the results to facilitate the use and interpretation of the ACRatio for athletes and coaches.

## 2.6 Increased injury risks at low ACRatio are expected due to bias

The ACRatio-injury risk model included in the IOC statement shows an increasing risk of injury at ACRatios below 0.8 (Figure 1). This suggests that performing less activity than one has prepared for in the previous weeks results in a higher risk of injury than performing the same level of activity. It also suggests that progressively lower amounts of activity result in further increases in injury risk. However, there is no biological theory that would support this hypothesis, as injury only occurs when the load exceeds the load capacity of the tissue [30]. This finding may be due the likelihood of sparse data bias, as mentioned above, as the number of participants who are injured at low ACRatios is expected to be quite small relative to higher ACRatios. However, the observed increase in injury risk associated with low ACRatios in Figure 1 can also be explained by bias due to its method of calculation.

Activity data has typically been categorized into weekly blocks from Monday through Sunday based on calendar time [4,7,8,10]. Consider the following example: two athletes have the same chronic workload. One athlete has planned for 2 hours of activity each day for the current week, while the other athlete has planned for 1 hour of activity each day. If the larger workload causes injury for Athlete 1 on Tuesday, they will have an acute workload of 4 hours. Meanwhile,

Athlete 2 completes the week without injury, with an acute workload of 7 hours. Athlete 1 will have a lower ACRatio than Athlete 2, despite performing more activity just prior to injury. Because athletes who get injured earlier in the week will have systematically lower acute workloads than athletes who complete the week without injury, the current analytical strategies suggested for the ACRatio will show an increased injury risk for low ratios even though there is no causal relationship.

There are three alternative methods of calculation that researchers have used in the literature. First, one could measure injury risk in the week subsequent to the acute workload [7]. Although measuring injury risk for the subsequent week circumvents low ACRatios being associated with high risks of injuries from athletes getting injured early in the week, it ignores variations in the current week that likely impact injury risk. For example, an athlete with a low ACRatio may have a sudden spike in activity in the subsequent week that causes injury. However, analyses would consider their low ACRatio from the previous weeks to be predictive of injury, rather than their spike in recent activity.

Second, one could measure the ACRatio of the current week with a two-week average workload from the current and prior week [7]. Calculating the acute workload as a two-week average faces the same problems as measuring injury risk in the current week – athletes who get injured early in the week will have a systematically lower ACRatio, although to a lesser extent.

Third, some have calculated acute and chronic loads as daily moving averages, where the acute load is the average daily workload in the last $X$ days (e.g. 7 days) and the chronic load is the average daily workload in the last $Y$ days (e.g. 28 days) [5]. The ACRatio for a particular day is the acute load divided by the chronic load, whereby the acute load excludes activity performed on that day. Therefore, injury risk is measured for the day subsequent to the last day included in the acute load. This method faces the same limitations as described above. Measuring injury risk for the subsequent day avoids the bias where low ACRatios are observed for injured athletes, but ignores any activity on that day that would have contributed to injury. On the other hand, including the current day's activity in the acute load calculation will result in lower ACRatios for athletes who get injured early in the day, albeit with reduced bias compared to the weekly calculations.

The underlying issue in the established methods is that injured and non-injured athletes are not matched on time. Athletes who get injured contribute less time during which they can add to their workload. One simple solution is to use a nested case-control design where one matches athletes based on exposure time, and "censors" the uninjured athletes at the calendar time that their matched athlete was injured. The ACRatio is then calculated for each matched pair and a matched analysis is conducted. Therefore, all activity performed by both athletes prior to the injury is included, and any activity performed after the injury by the uninjured athlete is excluded. To be truly matched on time, the athletes would need the same training schedule. This might be feasible for athletes on the same competitive team, but would be difficult for individual sports or across teams. Another similar option is to use a case-crossover design, where one includes only athletes who had an injury [31]. One then compares their ACRatio at a point in

1111time when they got injured with their ACRatio at another point in time when they were not injured. Both the matched nested case-control design and case-crossover study designs censor data from the uninjured athletes, discarding useful information. In addition, both designs suffer from some limitations in interpretation. The case-crossover design only informs the relationship between the ACRatio and injury in those who experience injury over the time period of observation. They do not necessarily provide information on the average causal relationship between the ACRatio and injury in the entire population, without additional assumptions. The nested-case control design suffers from a similar limitation, in that inference is limited to only those subjects who could have been matched to athletes with injuries, again without additional assumptions.

Although matching on time provides a solution when activity is recorded daily, many studies only gather weekly activity data for feasibility reasons. In these studies, an unbiased estimate of the acute load could be estimated with some reasonable assumptions if the "training schedule" is known. To illustrate, consider two athletes that have the same chronic workload. Athlete 1 plans for 2 hours of activity each day for the current week, but gets injured on Tuesday at the end of the workout, totalling 4 hours of activity. Athlete 2 plans for 1 hour of activity each day, and completes the week without injury, totalling 7 hours of activity. In this example, Athlete 1's acute load is 2/7 of their weekly workload (2/7 x 14 hours = 4 hours). Although Athlete 2 completed 100% of their acute load during the week, we could censor their activity at 2/7 of their weekly workload (2/7 x 7 hours = 2 hours). This simple calculation will work under the strong assumption that daily activity is equal throughout the week. Alternatively, if the planned training schedule is known and generally followed, one could simply use the planned workload as a proxy for the true workload. We strongly recommend that training schedules be recorded whenever possible, and used as a proxy for actual activity where reasonable.

## 2.7 The current ACRatio-injury risk model does not address unmeasured confounding

For most questions concerning activity workload and injury, there likely exists important confounding factors that were unmeasured or not measured well. One strategy for obtaining an unbiased estimate of an association between exposure and outcome in the presence of unmeasured confounding is to use an instrumental variable (IV) approach [32]. An instrument is a variable that causes the exposure, does not cause the outcome except through its effects on exposure, and is not caused by any confounders of the exposure and outcome relationship. The causal effect of the exposure on the outcome may be estimated using IV estimation methods such as 2-stage least squares regression, and will be unbiased provided the IV assumptions hold [33]. In the context of the ACRatio and injury, an athlete's proposed training schedule might be a valid instrument for their actual activity because it would only affect injury through the actual activity performed. If a coach is aware of factors that increase the risk of injury for one of their athletes, they might modify the training schedule for that athlete. These confounding factors can be conditioned on in the IV analysis in order to satisfy the assumption that there is no confounding between the IV and the outcome. When collecting activity and injury data, we recommend that investigators record any reasons for changes in training schedules that might be related to injury risk to allow for more advanced causal inference type analyses.

Another important analytical challenge for many questions concerning activity load and injury is that activity loads, and the ACRatio itself, are time-varying measures [34] that are also affected by other time-dependent variables. For instance, one's activity load may cause them to develop aches and pains that, while not considered injuries, may both predispose them to injury (affecting the outcome) and change their activity load in the following week (affecting the exposure). Because the ACRatio affects time-dependent variables that in turn affect the outcome, traditional logistic regression methods are not appropriate [34]. More advanced methods, such as g methods [35], should be explored. Such methods require understanding why athletes deviate from their training schedule, again highlighting the advantages of obtaining planned training schedules and deviations from them as part of the study protocol for questions related to activity and injury.

## 2.8 Difficulties applying the ACRatio to subsequent injuries

The ACRatio has been used to develop training recommendations to prevent injuries. In fact, many injuries occur in athletes who were previously injured. These subsequent injuries make up a large portion of injuries [36–38], and as such, using the ACRatio to guide their prevention may be of interest.

There are several considerations that are unique to subsequent injuries and limit the use of the regular ACRatio-injury risk model. First, previous research using the ACRatio has treated injuries in each athlete as independent events. An injury that occurs in each of 4 athletes should be analyzed differently than 4 injuries that occur in one athlete, as the latter is an outcome with recurrent events [39].

Second, injuries lead to changes in training schedules. Athletes often lose training time upon getting injured [40]. If an athlete returns to play but gets re-injured during the chronic load time frame, the ACRatio will be inflated due to a smaller denominator resulting from training lost to injury.

Third, injuries lead to changes in load capacity. Subsequent injuries may occur because athletes are not completely rehabilitated from injury. Training without having completely recovered from injury may put them at higher risk for injury than a healthy athlete, and they may experience injury at lower ACRatios. That is, a subsequent injury may not only be caused by workload, but by the previous under-rehabilitated injury, either separately or in combination. If the ACRatio decreases because of a previous under-rehabilitated injury, an unbiased analysis must account for this relationship when analyzing the effect of ACRatio on future injury.

For these reasons, the current ACRatio thresholds for minimizing injury risk are unlikely to hold when applied to subsequent injuries. A subsequent injury model needs to account for both time-varying confounders, and the extent of and recovery from the previous injury. Therefore, more complex analyses, such as g-methods [35], may be required to model ACRatios for athletes dealing with subsequent injuries.

14## 3. Conclusion

The ACRatio is a promising concept to predict the relationship between activity and injury risk. However, its current formulation suffers from several limitations. The ACRatio should exclude the acute load from the chronic load calculation to prevent spurious correlations. The EWMA method gives loads further in the past less weight but creates an "initial value" problem that must be addressed, and may still induce spurious correlations. Injury risk models developed using continuous ACRatios, with discretization following model selection, are likely to be more informative because there are no arbitrary thresholds or loss of data. To prevent bias in the ACRatio for injured athletes and to apply the ACRatio to recurrent injuries, new analytical strategies must be developed.

# Figures

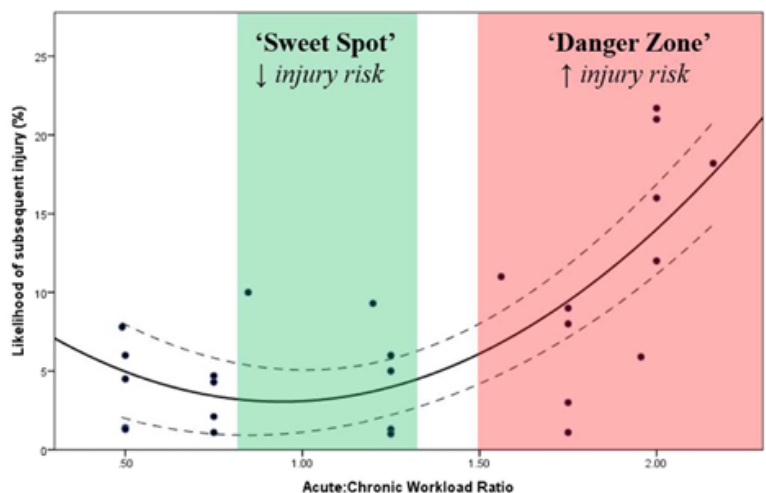

**Fig.1** Established relationship between the acute:chronic workload ratio and injury risk based on studies from 3 different sports. The green-shaded area covers acute:chronic workload ratios within a range of approximately 0.8–1.3, and represents a 'sweet spot' where injury risk is low. The red-shaded area covers acute:chronic workload ratios of 1.5 and above, and represents a 'danger zone' where injury risk is high [reproduced from [6]]

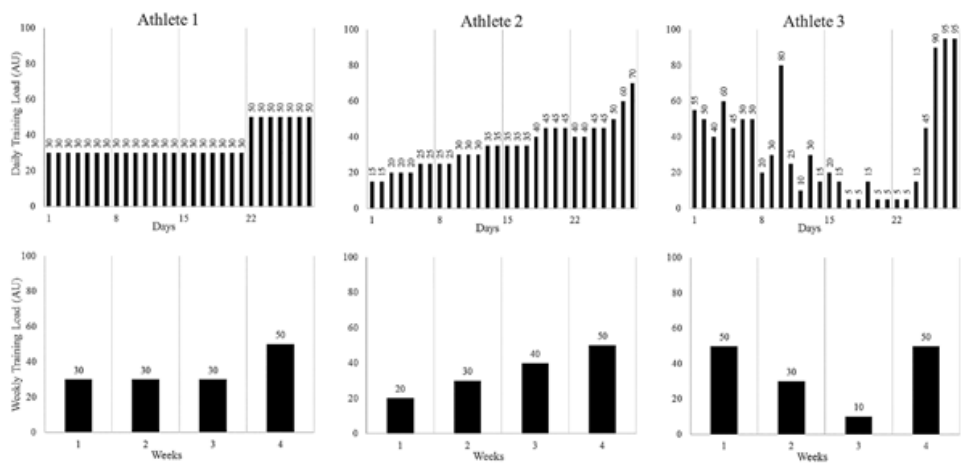

**Fig. 2** The daily training loads of three athletes are shown at the top and their weekly training loads are shown at the bottom. Each athlete has the same acute load (50 arbitrary units), chronic load (35 arbitrary units), and acute:chronic workload ratio (1.43) despite having different training schedules [reproduced from [17]]

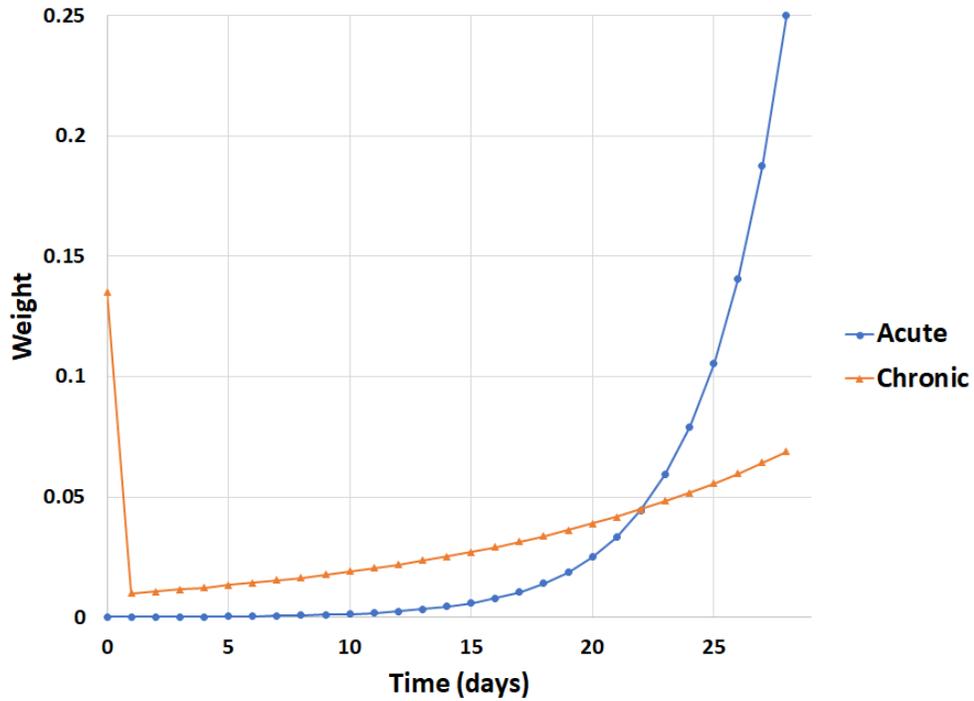

**Fig. 3** Weights for acute and chronic loads using the exponentially weighted moving average with 28 time points. The initial load takes on an inappropriately large weight for the chronic load, but its weight is negligible for the acute load

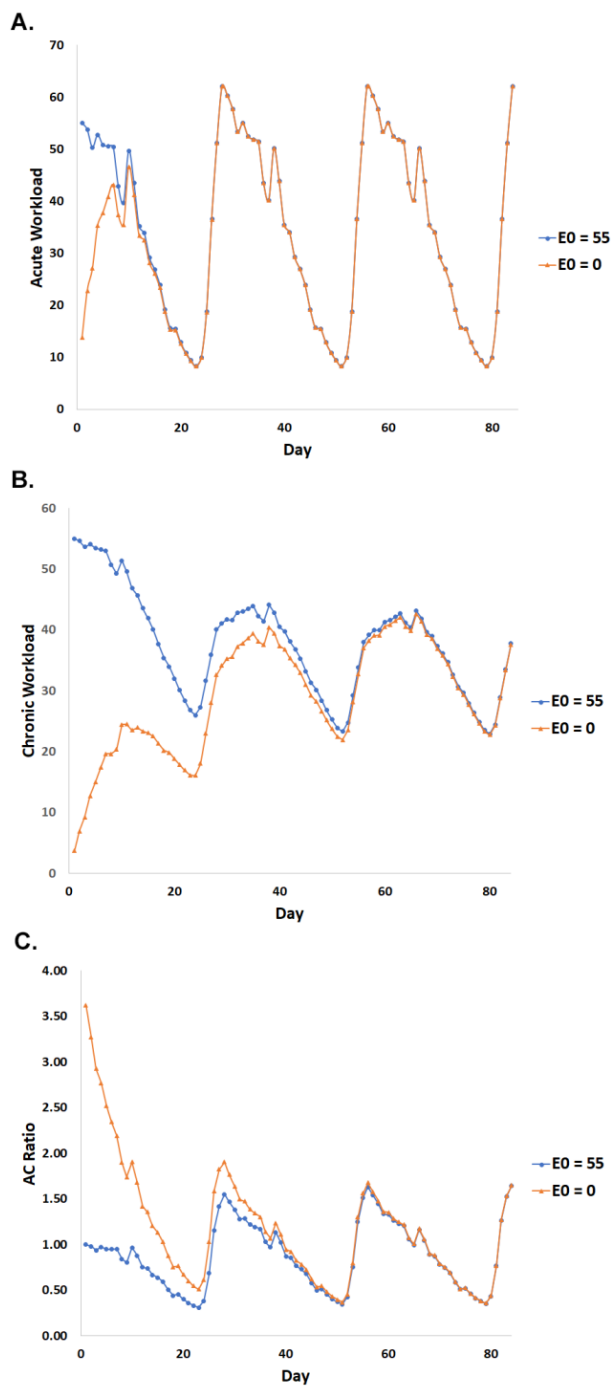

**Fig. 4** Acute and chronic loads calculated using the exponentially weighted moving average (EWMA) and their corresponding acute:chronic workload ratio (ACRatio) for two different initial values (E0). The two athletes follow the same training schedule with a load of 55 on day 1. **A.** EWMA-calculated acute load. The loads converge after 20 days **B.** EWMA-calculated chronic load. The loads converge after 55 days **C.** Acute:chronic workload ratio, calculated by dividing the acute load by the chronic load. The two acute:chronic workload ratios converge after 50 days

# Appendix 1

Using the coupled measure, an athlete or coach who wants to plan for the upcoming week, has to solve the following formula for acute workload:

$$acute\_workoad = \frac{max\_acceptable\_ACRatio * 3 * average_{3_{previous}}}{(4 - max\_acceptable\_ACRatio)} \quad \text{[eq A1.1]}$$

*Derivation of formula*

$$\frac{acute\_workload}{chronic\ workload} = max\_acceptable\_ACRatio \quad \text{[eq A1.2]}$$

$$\frac{acute\_workload}{\frac{(3*average_{3_{previous}} + acute\_workload)}{4}} = max\_acceptable\_ACRatio \quad \text{[eq A1.3]}$$

$$\frac{4*acute\_workoad}{(3*average_{3_{previous}} + acute\_workload)} = max\_acceptable\_ACRatio \quad \text{[eq A1.4]}$$

$$4 * acute\_workload = max\_acceptable\_ACRatio * (3 * average_{3_{previous}} + acute\_workload) \quad \text{[eq A1.5]}$$

$$4 * acute\_workload = max\_acceptable\_ACRatio * 3 * average_{3_{previous}} + max\_acceptable\_ACRatio * acute\_workload \quad \text{[eq A1.6]}$$

$$4 * acute\_workload - max\_acceptable\_ACRatio * acute\_workload = max\_acceptable\_ACRatio * 3 * average_{3_{previous}} \quad \text{[eq A1.7]}$$

$$acute\_workload * (4 - max\_acceptable\_ACRatio) = max\_acceptable\_ACRatio * 3 * average_{3_{previous}} \quad \text{[eq A1.8]}$$

$$acute\_workoad = \frac{max\_acceptable\_ACRatio * 3 * average_{3_{previous}}}{(4 - max\_acceptable\_ACRatio)} \quad \text{[eq A1.9]}$$

*Numerical Example*
Using the IOC consensus statement's maximum acceptable ACRatio of 1.3, and an athlete performing an average of 10 units of activity per week for the previous 3 weeks:

The coupled measure suggests the planned acute workload should not exceed:

$$acute_{workload} = \frac{1.3 * 3 * average_{3_{previous}}}{4 - 1.3} = \frac{3.9 * 10}{2.7} = 14.4\ units$$

The uncoupled measure will normally have a different acceptable upper limit than the coupled measure. Because the analyses for the uncoupled measure have not been conducted, for pedagogical purposes in this example, we consider the acceptable upper limit to also be 10 units.

In this case, the upcoming acute workload should not exceed 13 units [1.3*(previous chronic workload)]. Similar calculations would be required for the lower limits.

## Appendix 2

The exponentially weighted moving average (EWMA) is calculated as:
$$EWMA_{today} = \lambda \cdot Load_{today} + (1 - \lambda) \cdot EWMA_{yesterday}, \quad (A2.1)\ [18]$$

where $\lambda$ is the degree of decay and takes on a value between 0 and 1. When $\lambda = 1$, the EWMA is equal to $Load_{today}$, and the previous loads are not considered when calculating the average load. When $\lambda = 0$, the EWMA is equal to the initial load (i.e. the first recorded load) and is constant each day. Williams et al. chose $\lambda$ as

$$\lambda(N) = \frac{2}{N + 1}$$

where N is a time decay constant and equals 7 for the acute load and 28 for the chronic load [18]. Therefore the decay constant $\lambda(7)$ is $2/(7+1) = 0.25$ for the acute load and $\lambda(28) = 2/(28+1) = 0.068$ for the chronic load.

Denoting $EWMA = E$, $Load = L$, $today = t$, and $yesterday = t - 1$, for $t \geq 1$, equation A2.1 can be rewritten as

$$E_t = \lambda \cdot L_t + (1 - \lambda) \cdot E_t \quad (A2.2)$$

Recursively applying equation A2.2, we have

$$E_t = (1 - \lambda)^t E_0 + \lambda(1 - \lambda)^{t-1} L_1 + \lambda(1 - \lambda)^{t-2} L_2 + \cdots \\ + \lambda(1 - \lambda)^2 L_{t-2} + \lambda(1 - \lambda)^1 L_{t-1} + \lambda(1 - \lambda)^0 L_t \quad (A2.3)$$

where $E_0$ is the initial value. The weight applied to a specific day in the EWMA calculation is equal to $w_i = \lambda(1 - \lambda)^{t-i}$, where $t$ is the total number of days (e.g. 7 for the acute load), and $i$ is the number of days from the start day (e.g. 0 for the first day, 1 for the second day). The weight for the initial load that takes place on the first day of the calculation is equal to $w_0 = (1 - \lambda)^t$. Equation A2.3 can thus be rewritten as

$$E_t = w_0 E_0 + w_1 L_1 + w_2 L_2 + w_{t-2} L_{t-2} + w_{t-1} L_{t-1} + w_t L_t \quad (A2.4)$$

or more simply,

$$E_t = w_0 E_0 + \sum_{i=1}^{t} w_i L_i. \quad (A2.5)$$

Since $\lambda$ must be between 0 and 1, we have
$$w_1 < w_2 < \cdots < w_{t-1} < w_t, \quad (A2.6)$$

and we would like

$$w_0 < w_1. \quad (A2.7)$$

However, this last condition only applies when $\lambda > 0.5$ because:

$$w_0 = (1-\lambda)^t \quad (A2.8)$$
$$w_1 = \lambda(1-\lambda)^{t-1} \quad (A2.9)$$

Therefore, $w_0 < w_1$ when:

$$(1-\lambda)^t < \lambda(1-\lambda)^{t-1} \quad (A2.10)$$

$$1-\lambda < \lambda \quad (A2.11)$$

$$1/2 < \lambda \quad (A2.12)$$

To calculate the acute and chronic loads, the proposed $\lambda$'s are $\lambda(7) = 2/(7+1) = 0.25$ for the acute load and $\lambda(28) = 2/(28+1) = 0.068$ for the chronic load. These values are smaller than 0.5, and therefore $w_0 > w_1$. Therefore, the weight for the initial value will always be greater than the weight for the first load.

# Appendix 3

**Table 1.** Exponentially weighted moving average (EWMA) weights for the first two loads depending on the decay constant ($\lambda$) for 28 days of activity. $w_0$ denotes the initial load (day 1) and $w_1$ denotes the second load (day 2). The difference between weights increases as $\lambda$ gets closer to zero.

| $\lambda$ | $w_0$ | $w_1$ | Difference ($w_1$-$w_0$) |
|---|---|---|---|
| 0.500 | 0.000000004 | 0.000000004 | 0.000000000 |
| 0.475 | 0.000000015 | 0.000000013 | -0.000000001 |
| 0.450 | 0.000000054 | 0.000000044 | -0.000000010 |
| 0.425 | 0.000000187 | 0.000000138 | -0.000000049 |
| 0.400 | 0.000000614 | 0.000000409 | -0.000000205 |
| 0.375 | 0.000001926 | 0.000001156 | -0.000000770 |
| 0.350 | 0.000005775 | 0.000003110 | -0.000002666 |
| 0.325 | 0.000016615 | 0.000008000 | -0.000008615 |
| 0.300 | 0.000045999 | 0.000019714 | -0.000026285 |
| 0.275 | 0.000122875 | 0.000046608 | -0.000076267 |
| 0.250 | 0.000317479 | 0.000105826 | -0.000211653 |
| 0.225 | 0.000795147 | 0.000230849 | -0.000564298 |
| 0.200 | 0.001934281 | 0.000483570 | -0.001450711 |
| 0.175 | 0.004578367 | 0.000971169 | -0.003607198 |
| 0.150 | 0.010561605 | 0.001863813 | -0.008697792 |
| 0.125 | 0.023780747 | 0.003397250 | -0.020383497 |
| 0.100 | 0.052334763 | 0.005814974 | -0.046519790 |
| 0.075 | 0.112711575 | 0.009138776 | -0.103572798 |
| 0.050 | 0.237826885 | 0.012517204 | -0.225309681 |
| 0.025 | 0.492185981 | 0.012620153 | -0.479565828 |